\documentclass[eat,twocolumn]{jmlr}

\usepackage{tikz}
\usepackage{multirow}
\usetikzlibrary{positioning}
\usepackage{amsfonts}
\usepackage{amsmath}
\newcommand{\comment}[1]{}
\usepackage[english]{babel}
\usepackage{hyphenat}
\usepackage{longtable}% for long tables

\usepackage{booktabs}
\usepackage[load-configurations=version-1]{siunitx} % newer version
\theorembodyfont{\upshape}
\theoremheaderfont{\scshape}
\theorempostheader{:}
\theoremsep{\newline}

\jmlrvolume{}
\firstpageno{1}

\jmlryear{2022}
\jmlrworkshop{Machine Learning for Health (ML4H) 2022}

\title[HHMMs for Sleep Activity Recognition from Multi-Source Passively Sensed Data]{Heterogeneous Hidden Markov Models for Sleep Activity Recognition from Multi-Source Passively Sensed Data}

   \author{\Name{Fernando Moreno-Pino} \Email{fmoreno@tsc.uc3m.es}\\
   \Name{María Martínez-García} \Email{mariamartinezga@tsc.uc3m.es}\\
   \Name{Pablo M. Olmos} \Email{olmos@tsc.uc3m.es}\\
   \Name{Antonio Artés-Rodríguez} \Email{antonio@tsc.uc3m.es}\\
   \addr Signal Processing and Learning Group, Dept. of Signal Theory and Communications\\ Universidad Carlos III de Madrid, Spain}

\begin{document}

\maketitle
\begin{abstract}
Psychiatric patients' passive activity monitoring is crucial to detect behavioural shifts in real-time, comprising a tool that helps clinicians supervise patients' evolution over time and enhance the associated treatments' outcomes. Frequently, sleep disturbances and mental health deterioration are closely related, as mental health condition worsening regularly entails shifts in the patients' circadian rhythms. Therefore, Sleep Activity Recognition constitutes a behavioural marker to portray patients' activity cycles and to detect behavioural changes among them. Moreover, mobile passively sensed data captured from smartphones, thanks to these devices' ubiquity, constitute an excellent alternative to profile patients' biorhythm.
 In this work, we aim to identify major sleep episodes based on passively sensed data. To do so, a Heterogeneous Hidden Markov Model is proposed to model a discrete latent variable process associated with the Sleep Activity Recognition task in a self-supervised way.
We validate our results against sleep metrics reported by clinically tested wearables, proving the effectiveness of the proposed approach. 
\end{abstract}

\section{Introduction}
\label{sec:intro}

Mental illnesses affect as much as the 2\% of world's population \citep{WHO} and they are estimated to be the cause of about 14\% of global diseases \citep{prince2007no}. Mental disorders lead to quality of life deterioration along with social disabilities that burden patients' professional and social lives \citep{rapaport2005quality}. Therefore, monitoring patients that suffer from mental illnesses is vital to detect trend changes that may indicate behavioural shifts. 
%Mental disorders, which affect as much as 2\% of world's population \citep{WHO} and are estimated to be the cause of about 14\% of global disease \citep{prince2007no}, lead to quality of live deterioration along with social disabilities that burden patients' professional and social life \citep{rapaport2005quality}.
%Monitoring patients who suffer from mental illnesses is vital to detect trend changes that may indicate behavioral shifts. 
Historically, psychiatric patients' mental state monitoring has been assessed via questionnaires, patients' reports, and periodic appointments, which entail long periods before detecting any possible risk. Recent technological advances allow real-time patients monitoring with tools like the smartphone-based active Ecological Momentary Assessment (EMA) \citep{porras2020smartphone}. 
Nevertheless, collected data via electronic EMA suffer from patients' subjective insight and presumes patients' ability and interest to fill in the required information \citep{fischer2021possibilities}.
Objective assessment methods can be achieved through passive data collected from patients' smartphones, using information as the visited locations via GPS, smartphone and social media interaction, etc. The ubiquity of smartphone devices allows Human Activity Recognition (HAR) methods \citep{jobanputra2019human, rios2020hidden} to track patients in a non-invasive way, creating a personal profile for each user. To do so, the variety of smartphones' available sensors must be merged into relevant indicators of patients' daily life. These indicators constitute biomarkers that represent patients' state in real-time and can be used to profile their evolution, allowing real-time monitoring and detection of behavioural shifts.
Recently, several studies have identified relations between mental health condition worsening and sleep disruption in psychiatric patients \citep{tseng2019sleep, littlewood2019short, freeman2020sleep}. Considering the available data through patients' smartphones, sleep estimation arises as a natural biomarker to integrate smartphones' data into a single behaviour metric that characterizes each individual's circadian rhythm over time. Furthermore, a wide range of studies have reported that wearable devices show high reliability while detecting asleep-awake states in different populations, consistently reporting good estimators of the sleep activity \citep{de2018validation,chinoy2021performance}, which allows them to be used for validating predictions performed by Sleep Activity Recognition (SAR) models, as they can assess sleep quality during long time windows.

%In this work, we aim to create a Sleep Activity Recognition method, providing clinicians a monitoring tool for daily behavioral assessment.

%Once this patient characterization has been completed, changes in the initial trends may represent useful information to understand patients' evolution along time, allowing for preventive detection on significant behaviour changes.

\vspace{-0.2cm}
\section{Sleep Measurement Techniques}
\label{sec:sleep_measurement}

This work aims to identify major sleep episodes based on passively sensed behavioural data, obtaining objective behaviour markers that could be used to recognise major disturbances in psychiatric patients. 
Among traditional sleep assessment methods, Polysomnography (PSG) stands as the most reliable technique. Nonetheless, PSG requires medical assistance for its application and sleep studies need to be recorded in an ambulatory setting \citep{smith2020actigraph}. Besides, it is an expensive procedure that does not provide a continuous evaluation of patients' sleep as it is required to assess their behaviour evolution.
Sleep questionnaires and diaries arise as a natural alternative when trying to monitor sleep evolution constantly. Regardless, we discard them as they do not express an objective sleep evaluation \citep{hennig2017predicts}.
In these conditions, wearable devices appear as an excellent alternative: they are not expensive and they collect daily information in a non-intrusive way. Several studies have shown that, even when they are not sufficiently accurate to be used in clinical settings, mainly due to poor specificity and low accuracy, they can be relied on for detecting asleep-awake states \citep{moreno2019validation}. Nonetheless, the percentage of users who possess these devices is minimal, implying that we cannot rely on wearable devices to assess patients' sleep continuously.
Therefore, smartphones constitute the most reliable source of data: most patients have one, they do not require an additional investment, and most people carry them along most of the day; the ubiquity of these devices makes them the perfect passive assessment. However, mobile health data that can be collected this way usually entails several problems: missing data is quite common, signals are typically corrupted by noise, the different number of sensors that smartphones use imply data heterogeneity, and main sleep trends must be detached from false positives while making the predictions. Therefore, the proposed model must be capable of dealing with these problems.
Several frameworks are available in the literature to obtain sleep estimates from smartphone data \citep{gautam2015smartphone, eichi2021dpsleep}, as well as certain HMM implementations that handle mixed discrete and continuous observations \citep{epaillard2015hybrid, missaoui2013multi}. Anyhow, none of these models can simultaneously deal with all the complex requirements previously mentioned, making them not suitable for creating behavioural markers in the considered scenario. 
Furthermore, the literature lacks models validated with clinical patients and tested in large samples \citep{aledavood2019smartphone}. For these reasons, we propose our own method for the SAR problem, detailed in Section \ref{sec:sar}.

%\section{Passively Sensing behavioral Mobile Data ó Passively Sensing Mobile Sensor Data}
\vspace{-0.2cm}
\section{Passively Sensing Mobile Data}\label{sec:dataset}

\begin{figure*}[]
\vspace{-0.4cm}
\centering
    {\includegraphics[width=13cm, height=5.3cm]{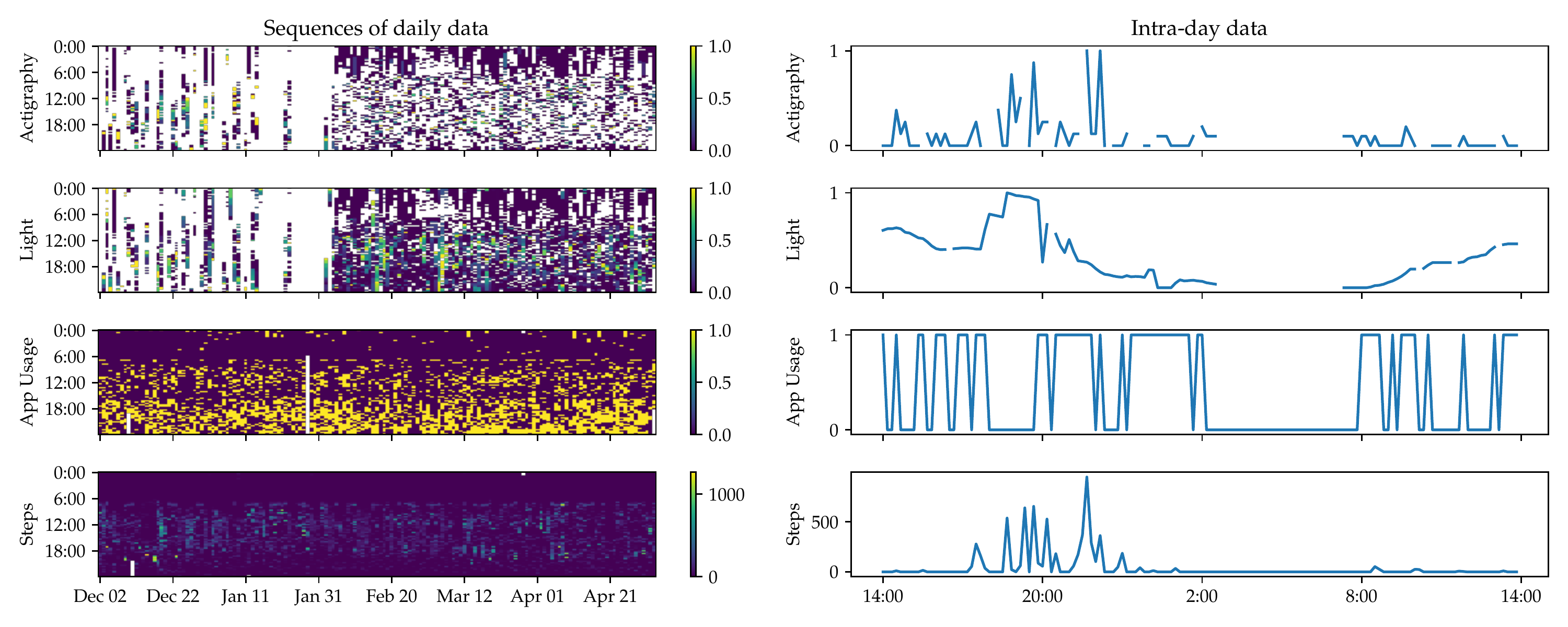}}
    %{\includegraphics[width=8cm]{figures/tmp/user_workplace_5_daily_repr.png}}
    %\vspace{-0.35cm}
    %\subfloat[Test-set features with 10\% and 20\% abstention rate after feature selection.]{
    %{\includegraphics[width=5.0cm]{./images/plot_of_0.1sk}}
    %{\includegraphics[width=5.0cm]{./images/plot_of_0.2sk}}}
\caption{Patient's passively sensed data. The left figure shows five months of data, the right figure displays intra-day features' values.}
\label{fig:dataset}
%\vspace{-0.35cm}
\vspace{-0.6cm}
\end{figure*}

In this section, we aim to specify the details of the dataset used. As previously stated, smartphones' data constitute a tool for patient monitoring due to their omnipresence, reporting objective measures of patients' circadian rhythm that can be used as indicators for detecting behavioural shifts. A data gathering system \citep{eB2} was used to collect passively sensed mobile data from patients' smartphones under clinicians' supervision. 
The available population consisted of a set of 57 patients who suffered from different mental health conditions. For 10 of those patients, 386 days of data with associated wearables' sleep metrics were available. Data from this subset of patients was exclusively used for validation purposes and did not take part during the training process. Demographic information for this subset was not available.
 The rest 47 patients, 38 females and 9 males, had an average age of 43 $\pm$ 15 years. For this second subset of patients, 647 days of data were collected. Thus, considering both sets, 1033 days of data were available for this study, with an average of 15 $\pm$ 24 days of data per user. We should clarify that train/test split was done solely depending on the availability of wearable devices' data: those patients who possessed wearable devices were not used during the training phase. Instead, their corresponding data was kept for testing purposes, as wearable devices’ reported sleep metrics allow us to assess the quality of the model’s predictions. This is possible as the proposed model operates in an unsupervised way, which entails that no labelled data is required during training, allowing us to use all wearable devices’ data for evaluation purposes.
Data acquisition was conducted periodically, collecting data from different sensors, i.e. actigraphy, light, smartphone usage (a binarized feature that indicates if the patient is using or not the smartphone), and pre-computed metrics, i.e. the daily amount of steps, which we binarize for each time-slot. Data's granularity  was 10 minutes; hence 144 observations per day were available in absence of missing data. Table \ref{tab:data} summarizes the smartphones' used sensors and Figure \ref{fig:dataset} shows the collected data's structure for one of the patients.
The models  developed to extract relevant sleep trends from this real-world data (incomplete, noisy, and heterogeneous) need to consider its structure, as it encloses different challenges while building optimal methods capable of dealing with these problems.

\comment{
\begin{figure*}[t]
\centering
\includegraphics[width=0.4\textwidth]{example-image-golden} % Reduce the figure size so that it is slightly narrower than the column.
\caption{Here we will show a couple of figures to illustrate the dataset strucutre}
\label{fig1}
\end{figure*}
}

\begin{table}[]
\fontsize{10.0}{12}\selectfont
\centering
\begin{tabular}{ccc}
\specialrule{.1em}{.05em}{.05em}
\textbf{Feature} & \textbf{Domain} & \textbf{\% Missing values} \\ \specialrule{.1em}{.05em}{.05em}
Actigraphy                      & $\mathbb{R}$             & 26.27 \%                   \\
Light                           & $\mathbb{R}$               & 59.22 \%                  \\
Steps                           & $\mathbb{Z}$               & 21.98 \%                      \\
App usage                       & $\mathbb{Z}_{2}=\{0,1\}$     & 5.33 \%                       \\ \specialrule{.1em}{.05em}{.05em}
\end{tabular}
\caption{Passively sensing mobile data features used and missing observations percentage per feature.}
\label{tab:data}
\vspace{-1.0cm}
\end{table}

%\section{Passively Sensing behavioral Mobile Data ó Passively Sensing Mobile Sensor Data}
\vspace{-0.5cm}
\section{Sleep Activity Recognition}
\label{sec:sar}

%$\mathbb{N}, \mathbf{N}$

Passively sensing mobile data's characteristics, enumerated in Sections \ref{sec:sleep_measurement} and \ref{sec:dataset}, demarcates the features that we demand from the proposed model. Firstly, it needs to be able to manage data heterogeneity, handling gaussian and categorical observations at the same time. Secondly, it needs to perform inference on missing data, critical while dealing with healthcare datasets. Finally, it needs to allow semi-supervised training, granting users the possibility to fix certain observation probabilities, therefore manually designing certain aspects of the model’s behaviour. E.g., fixing to zero the probability of observing smartphone usage for those hidden states that we aim to interpret as sleeping activity. %I.e., we could fix to zero the probability of observing smartphone usage while sleeping for those hidden states that we aim to interpret as sleeping activity, while inferring the observation probabilities for the rest of the features.
Fullfilling these requirements, we present a model easy to train and with a low number of parameters, which allows it to be implemented in a production setting and to be used by clinicians as a tool to monitor patients’ circadian rhythm, unveiling a powerful mechanism to detect behavioural changes.

To merge the heterogeneous input features that define the collected data into a SAR indicator, capable of detecting different activity levels along the day and identifying sleep episodes among them, we require the proposed model to be capable of projecting the high dimensional input features into a lower dimensional space. To do so, latent variable models arise as a natural alternative \citep{bishop1998latent}. Considering the time dependence of our data, Hidden Markov Models (HMMs) \citep{rabiner1989tutorial} constitute a natural way of modelling a discrete latent process associated with the SAR task.
HMMs are generative models that assume that the phenomena being modelled is a Markov process \citep{ethier2009markov} with associated hidden states. HMMs' objective is to learn about these hidden states sequence, denoted $S=\left\{s_{1}, s_{2}, \ldots, s_{T}: s_{t} \in 1, \ldots, I\right\}$, with $t=\left\{1, 2, \ldots, T\right\} \in  \mathbb{N}$, by means of the generated observations, $Y=\left\{\mathbf{y}_{1}, \mathbf{y}_{2}, \ldots, \mathbf{y}_{T}: \mathbf{y}_{t} \in \mathbb{R}^{M}\right\}$. To do so, they use an observation model $p(\mathbf{y}_t | s_t)$. Therefore, each observation $\mathbf{y}_t$ depends exclusively on its associated state $s_t$.
Here, we aim to use this hidden states sequence $S$ as the low dimensional representation of the input features. Therefore, each possible state of the HMM is associated with different activity levels, being  sleep episodes one of them.

\begin{figure}[]
%\vspace{-0.2cm}
\centering
\begin{tikzpicture}[scale=0.9, transform shape]
\tikzstyle{main}=[circle, minimum size = 10mm, thick, draw =black!80, node distance = 10mm]
\tikzstyle{connect}=[-latex, thick]
\tikzstyle{box}=[rectangle, draw=black!100]
  \node[main] (S1) [] {$s_{t-1}$};
  \node[main] (S2) [right=of S1] {$s_{t }$};
  \node[main] (S3) [right=of S2] {$s_{t+1}$};
  \node[main] (St) [right=of S3] {$s_{T}$};
  
  \node[main,fill=black!10] (O1) [below=of S1] {$y_{t-1}$};
  \node[main,fill=black!10] (O2) [right=of O1,below=of S2] {$y_{t}$};
  \node[main,fill=black!10] (O3) [right=of O2,below=of S3] {$y_{t+1}$};
  \node[main,fill=black!10] (Ot) [right=of O3,below=of St] {$y_{T}$};
  
  \node [main,fill=black!10] (L1) at (1,-1.2) {$l_{t-1}$};
  \node [main,fill=black!10] (L2) at (3,-1.2) {$l_{t}$};
  \node [main,fill=black!10] (L3) at (5,-1.2) {$l_{t+1}$};
  \node [main,fill=black!10] (Lt) at (7,-1.2) {$l_{T}$};

  \path (S3) -- node[auto=false]{\ldots} (St);
  \path (S1) edge [connect] (S2)
        (S2) edge [connect] (S3)
        (S3) -- node[auto=false]{\ldots} (St);

  \path (S1) edge [connect] (O1);
  \path (S2) edge [connect] (O2);
  \path (S3) edge [connect] (O3);
  \path (St) edge [connect] (Ot);
  \path (S1) edge [connect] (L1);
  \path (S2) edge [connect] (L2);
  \path (S3) edge [connect] (L3);
  \path (St) edge [connect] (Lt);

  \draw[dashed]  [below=of S1,above=of O1];
\end{tikzpicture}
\caption{Heterogeneous HMM architecture. Gray represents observed data.}
\label{hhmm}
\vspace{-0.4cm}
\end{figure}
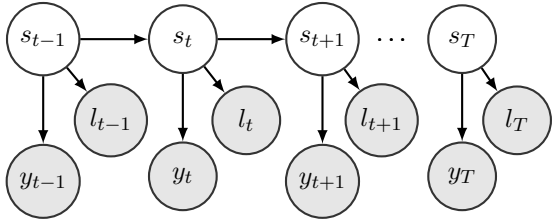

Nevertheless, standard HMM implementations use Multinomial and Gaussian observation models, depending on the probability distribution chosen to model the emission probabilities $p(\mathbf{y}_t | s_t)$.
Due to the heterogeneity of the collected data, neither of these options would fit our requirements. While some of the used features, like the actigraphy and light sensors, can be naturally modelled via Gaussian distributions, others, as the app usage and steps count (binarized and converted into an movement indicator), are better modelled with a Bernoulli distribution.
To manage the heterogeneity of our input sources, we developed a variation of ordinary HMMs which we refer to as Heterogeneous Hidden Markov Model (HHMM) \footnote{\tiny{https://github.com/fmorenopino/HeterogeneousHMM}} \citep{moreno2022pyhhmm}. The proposed HHMM can simultaneously manage categorical and continuous data  thanks to its observation model, visible in Fig. \ref{hhmm}, which is responsible for jointly modelling the continuous and discrete observations, $\mathbf{y}_t$ and $\mathbf{l}_t$. The joint distribution of the proposed model is stated in Eq. \ref{joint_main}, where $N \in  \mathbb{Z}$ represents the number of sequences. More information about the HHMM, its parameters inference procedure, and how it performs inference on missing data can be found in the Appendixes \ref{subsec:inference} and \ref{subsec:missing}.

\begin{equation}
\footnotesize
\begin{aligned}
    p(S, Y, L) =& \prod_{n=1}^{N}\left(p\left(s_{1}^{n}\right) \prod_{t=2}^{T_{n}}  p\left(s_{t}^{n} | s_{t-1}^{n}\right)\right) \\
    &\left(\prod_{t=1}^{T_{n}} p\left(\mathbf{y}_{t}^{n} | s_{t}^{n}\right)\right)\left(\prod_{t=1}^{T_{n}} p\left(l_{t}^{n} | s_{t}^{n}\right)\right)
\end{aligned}
\label{joint_main}
\vspace{-0.3cm}
\end{equation}

\vspace{-0.1cm}
\section{Experiments}
\label{sec:experiments}
We will now train the proposed model and evaluate its performance in comparison with different baseline methods. We should remark that because of the dataset used, where just a relatively small portion of data has the corresponding labels indicating the sleep activity of the patients, as stated in Section \ref{sec:dataset}, it is not feasible to train supervised methods due to this lack of labelled data.

HHMM's parameters are trained via the Baum-Welch algorithm using a subset the 47 patients with 647 sequences of daily data. The remainder 10 patients are retained for model evaluation purposes. The optimal number of hidden states is selected using two different model selection methods, the Bayesian Information Criteria (BIC) and the Akaike Information Criterion (AIC). For further details regarding the selection criteria for the optimal number of hidden states, as well as its projection into a binary output (awake or asleep), we refer the reader to Appendix \ref{sec:appendix_experiments}.
\comment{
*LARGO: no voy a hablar aquí de semi-supervised training
The training process with $N=6$ was performed in a semi-supervised manner. To do so, certain discrete emission probabilities were fixed in order to ban the model from accessing specific states if the smartphone was being used. Hence, we are manually associating certain hidden states to low activity periods. HHMM's predictions were then binarized into awake/asleep categories.
}
Once the optimal model's parameters are obtained, we proceed to evaluate the accuracy of the model in the task of detecting sleep episodes. To do so, we use the subset of 10 patients with 386 days of data for whom associated wearables' sleep metrics were available, reporting the sleep state of the patients as awake, light sleep, deep sleep or REM phase. We binarize this output into two categories: awake or asleep, so we can directly compare them with HHMM's final binary predictions.

\begin{table*}[]
\fontsize{10.0}{15}\selectfont
\centering
\begin{tabular}{ccccc}
\specialrule{.1em}{.05em}{.05em}
\multicolumn{2}{c}{\textbf{Method}}                                                                      & \textbf{Accuracy}                & \textbf{Specificity}             & \textbf{Sensitivity}             \\ \specialrule{.1em}{.05em}{.05em}
\multicolumn{2}{c}{\textbf{\begin{tabular}[c]{@{}c@{}}Random Classifier (Most Frequent)\end{tabular}}} & $0.6798 \pm 0.043$                   & $0.9973 \pm 0.0510$                  & $0.0026 \pm 0.0511$                  \\ \hline
\multicolumn{2}{c}{\textbf{\begin{tabular}[c]{@{}c@{}}Random Classifier (Uniform)\end{tabular}}}       & $0.4967 \pm 0.0426$                  & $0.5011 \pm 0.0502$                  & $0.4983 \pm 0.0762$                  \\ \hline
\multirow{2}{*}{\textbf{K-means}}                     & \textbf{Imputing Method 1}                     & $0.7393 \pm 0.1156$                  & $0.6338 \pm 0.1651$                  & $0.9641 \pm 0.0324$                  \\
                                                      & \textbf{Imputing Method 2}                     & $0.7386 \pm 0.1167$                  & $0.6356 \pm 0.1672$                  & $0.9587 \pm 0.0767$                  \\ \hline
\multirow{2}{*}{\textbf{GMM}}                         & \textbf{Imputing Method 1}                     & $0.5126 \pm 0.1327$                  & $0.2968 \pm 0.1926$                  & $0.9717 \pm 0.0416$                  \\
                                                      & \textbf{Imputing Method 2}                     & $0.7386 \pm 0.1167$                  & $0.6356 \pm 0.1672$                  & $0.9587 \pm 0.0767$                  \\ \hline
\multicolumn{2}{c}{\textbf{HHMM}}                                                                        & $0.8780 \pm 0.0804$ & $0.9052 \pm 0.0779$ & $0.8207 \pm 0.2382$ \\ \specialrule{.1em}{.05em}{.05em}
\end{tabular}
\caption{Experiments results.}
\label{tab:results}
\vspace{-0.2cm}
\end{table*}

In order to illustrate HHMM performance, we include two baseline models for comparison purposes. The first of them is a K-means algorithm that cluster each datapoint as asleep/awake. Two variations of this K-means model are considered, each of them uses different missing data imputation techniques in order to cluster each datapoint. The first imputation method (\textit{Imputation Method 1} in Table \ref{tab:results}) uses the mean to infer the missing values on the continuous features of the dataset, and it sets to zero the missing observations on the binary features. The second imputation method (\textit{Imputation Method 2} in Table \ref{tab:results}) still infers continuous missing values using the mean, but it substitutes missing observations on the binary features with the most common values (per sequence). 
Besides the K-means algorithm, the second baseline considered is a Gaussian Mixture Model (GMM), which also uses the two previously explained imputation techniques to deal with missing values. Furthermore, besides these two baselines, two dummy classifiers are tested on the labelled dataset. The first of these dummy classifiers follows a “most frequent” approach, always returning the most frequent class label. The second one is “uniform”, as it generates predictions uniformly at random from the list of unique classes observed. The inclusion of these two dummy classifiers provides a quantification of the label distribution of the dataset, which helps to understand class imbalances, as the portion of time the patients are asleep is considerable smaller than the time they are awake.

As done with the HHMM, we train the unsupervised baseline methods in the 47 patients within the training set. The trained methods are evaluated then in the cohort of 10 patients with associated labelled wearables’ data. The results reported by these baselines, together with HHMM's performance, are collected in Table \ref{tab:results}. As these results show, no baseline methods report better performance than the proposed algorithm in terms of accuracy. Regarding specificity and sensitivity, some of the utilised baselines report better results if we consider them separately, due to class imbalances, but the HHMM report the best trade-off between both metrics.
 For illustrative examples of the HHMM's predictions, we refer the reader to Figures \ref{fig:predictions1} and \ref{fig:predictions2} of the Appendix \ref{sec:appendix_experiments}, which show two examples of the HHMM's reported predictions for different patients.

\vspace{-0.3cm}
\section{Conclusion}
\label{sec:conclusion}

We introduce a new approach that performs Sleep Activity Recognition from passively sensed data, providing an unsupervised method that can operate in realistic scenarios and produces sleep activity indicators. Accordingly, the model comprises a tool that can be used by clinicians to monitor patients’ circadian rhythm.

A discrete latent variable model with temporal dependences is used to monitor patients' activity cycles, integrating smartphones’ data  into  a  single  behaviour  metric. The proposed model can deal with passively sensed data associated problems: corrupted signals, heterogeneity, and missing values. We validated our results against wearables' reported sleep metrics, proving the effectiveness of the proposed approach.
Therefore, Heterogeneous Hidden Markov Models comprise a powerful tool to supervise patient's evolution in real-time, providing a behavioural marker to portray patients' activity cycles and to detect behavioural changes.
\comment{

\section{Acknowledgments}

This work has been supported by Spanish government Ministerio de Ciencia, Innovación y Universidades under grants FPU18/00470, TEC2017-92552-EXP and RTI2018-099655-B-100, by Comunidad de Madrid under grants IND2017/TIC-7618, IND2018/TIC-9649, IND2020/TIC-17372,  and Y2018/TCS-4705, by BBVA Foundation under the Deep-DARWiN project, and by the European Union (FEDER) and the European Research Council (ERC) through the European Union’s Horizon 2020 research and innovation program under Grant 714161.
%\vspace{.2em}
% Use \bibliography{yourbibfile} instead or the References section will not appear in your paper
%\nobibliography{aaai22}
}

%\acks{Acknowledgements go here.}

\bibliography{pmlr-sample}

\appendix
\comment{
\section{Heterogeneous-Hidden Markov Model}
\label{subsec:hhmm}

The proposed Heterogeneous Hidden Markov Model (HHMM) can simultaneously manage categorical and continuous data  thanks to its observation model, which is visible in Section \ref{sec:sar}. This observation model is responsible of jointly modeling the continuous and discrete observations, $\mathbf{y}_t$ and $\mathbf{l}_t$. 

HHMMs can be fully characterized via the hidden states sequence, $S$; the continuous sequence observations, $Y$; its associated continuous observations emission probabilities, $\mathbf{B}=\left\{b_{i}: p_{b_{i}}\left(\mathbf{y}_{t}\right)=p\left(\mathbf{y}_{t} \mid s_{t}=i\right)\right\}$; the discrete sequence observations, $L=\left\{l_{1}, l_{2}, \ldots, l_{T}: l_{t} \in 1, \ldots, J\right\}$; its associated discrete observations emission probabilities, $\mathbf{D}=\left\{d_{i m}: d_{i m}=P\left(l_{t}=m \mid s_{t}=i\right)\right\}$; the state transition probabilities, $\mathbf{A}=\left\{a_{i j}: a_{i j}=p\left(s_{t+1}=j \mid s_{t}=i\right)\right\}$; and the initial state probability distribution, $\pi=\left\{\pi_{i}: \pi_{i}=p\left(s_{1}=i\right)\right\}$.
}

\section{Parameters Inference}
\label{subsec:inference}

Heterogeneous Hidden Markov Models, as classic HMMs, present three inference problems that must be solved to deliver an useful application.

The first of these problems is related with obtaining the probability of the observed variables, $\mathbf{Y}$ and $\mathbf{L}$, given the model parameters, $\theta=\{\mathbf{A}, \mathbf{B}, \mathbf{D}, \mathbf{\pi}\}$, for the considered time-steps $t=\left\{1, 2, \ldots, T\right\}$, i.e., to calculate the probability $p(\mathbf{Y}, \mathbf{L} | \theta)$. 

The second problem consists of determining the optimal hidden states path that better explains the observed data. This can be achieved by using the Forward-Backward algorithm, calculating $p(s_t | \mathbf{y}_t, \mathbf{l}_t)$ each time-step, or through the Viterbi algorithm, which maximizes the probability of the hidden states sequences by considering all time-steps $t=\left\{1, 2, \ldots, T\right\}$, i.e., calculating $p(S | \mathbf{y}_t, \mathbf{l}_t)$. 

Finally, the third problem consists of determining the optimal parameters $\theta$ that maximize the conditional probability $p(\mathbf{Y}, \mathbf{L} | \theta)$. The Baum-Welch algorithm, a special case of the Expectation-Maximization (EM) algorithm, can be used for obtaining this optimal parametrization. The joint distribution required for this third task is expressed in Eq. \ref{joint}.
For further information regarding these three problems and their solution, we refer the readers to \cite{rabiner1989tutorial}.
\begin{equation}
\footnotesize
\begin{aligned}
    p(S, Y, L) =& \prod_{n=1}^{N}\left(p\left(s_{1}^{n}\right) \prod_{t=2}^{T_{n}}  p\left(s_{t}^{n} | s_{t-1}^{n}\right)\right) \\
    &\left(\prod_{t=1}^{T_{n}} p\left(\mathbf{y}_{t}^{n} | s_{t}^{n}\right)\right)\left(\prod_{t=1}^{T_{n}} p\left(l_{t}^{n} | s_{t}^{n}\right)\right)
\end{aligned}
\label{joint}
\end{equation}

\comment{

*LARGO: no voy a hablar aquí de semi-supervised training

Lastly, we should mention that the proposed method can be trained in a semi-supervised manner, fixing discrete observations emission probabilities. When the model's parameters are trained in a semi-supervised way, discrete observations are refereed to as labels.
This semi-supervised way of training allows a guided learning process that improves model's parameters interpretability as certain states are associated with particular values of the labels.
}

\section{Missing Data Inference}
\label{subsec:missing}

As Table \ref{tab:data} proves, missing data is relatively common on passively sensing data. These missing observations can be either full missing, where none of the sensors hold a value, or partially missing, where specific sensors exhibit missing values. The model differentiates these two different scenarios and handles both of them through modifications in the Baum-Welch algorithm.%The first case can be solved by sampling the missing values from the posterior distribution. In the second case, when partial missing data appears, we must infer the missing values using the marginal distributions \citep{murphy2012machine} with respect to the observable features. 

%The first scenario, partial missing data, entails that among the features observed some of them exhibit a missing value, and supposing for simplicity's sake two jointly gaussian features at each time instant $\mathbf{x}=\left(\mathbf{x}_{1}, \mathbf{x}_{2}\right)$, with means:

The first scenario, full missing data, where all the features are non observable at a specific time instant, is straigh-forward solved by the Baum-Welch algorithm, which substitutes the missing values of each feature with its average per each hidden state.

In order to illustrate the second scenario, where partial missing data is present among the observed features, we assume for simplicity's sake two jointly gaussian features at each time instant, $\mathbf{x}=\left(\mathbf{x}_{1}, \mathbf{x}_{2}\right)$, with means:
    \begin{equation}
    \begin{aligned}
    \boldsymbol{\mu}&=\left(\begin{array}{l}
    \boldsymbol{\mu}_{1} \\
    \boldsymbol{\mu}_{2}
    \end{array}\right),
    \end{aligned}
    \label{eq:missing_parameters_1}
    \end{equation}
and covariance matrix:
    \begin{equation}
    \begin{aligned}
    \boldsymbol{\Sigma}=\left(\begin{array}{ll}
    \boldsymbol{\Sigma}_{11}  \boldsymbol{\Sigma}_{12} \\
    \boldsymbol{\Sigma}_{21}  \boldsymbol{\Sigma}_{22}
    \end{array}\right),
    \end{aligned}
    \label{eq:missing_parameters_2}
    \end{equation}
The inverse of this covariance matrix can be defined as follows:
    \begin{equation}
    \begin{aligned}
    &\boldsymbol{\Lambda}=\boldsymbol{\Sigma}^{-1}=\left(\begin{array}{ll}
    \mathbf{\Lambda}_{11} & \mathbf{\Lambda}_{12} \\
    \boldsymbol{\Lambda}_{21} & \boldsymbol{\Lambda}_{22}
    \end{array}\right),
    \end{aligned}
    \label{eq:missing_parameters_3}
    \end{equation}
\comment{
    \begin{equation}
    \begin{aligned}
    \boldsymbol{\mu}&=\left(\begin{array}{l}
    \boldsymbol{\mu}_{1} \\
    \boldsymbol{\mu}_{2}
    \end{array}\right), \boldsymbol{\Sigma}=\left(\begin{array}{ll}
    \boldsymbol{\Sigma}_{11}  \boldsymbol{\Sigma}_{12} \\
    \boldsymbol{\Sigma}_{21}  \boldsymbol{\Sigma}_{22}
    \end{array}\right), \\
    &\boldsymbol{\Lambda}=\boldsymbol{\Sigma}^{-1}=\left(\begin{array}{ll}
    \mathbf{\Lambda}_{11} & \mathbf{\Lambda}_{12} \\
    \boldsymbol{\Lambda}_{21} & \boldsymbol{\Lambda}_{22}
    \end{array}\right)
    \end{aligned}
    \label{eq:missing_parameters}
    \end{equation}
}
\comment{
\begin{equation}
\small
\boldsymbol{\mu}=\begin{pmatrix}
\boldsymbol{\mu}_{1}\\
\boldsymbol{\mu}_{2}
\end{pmatrix}, \boldsymbol{\Sigma}=\begin{pmatrix}
\boldsymbol{\Sigma}_{11}  \boldsymbol{\Sigma}_{12} \\
\boldsymbol{\Sigma}_{21}  \boldsymbol{\Sigma}_{22}
\end{pmatrix}, \boldsymbol{\Lambda}=\boldsymbol{\Sigma}^{-1}=\begin{pmatrix}
\mathbf{\Lambda}_{11} & \mathbf{\Lambda}_{12} \\
\boldsymbol{\Lambda}_{21} & \boldsymbol{\Lambda}_{22}
\end{pmatrix}
\hspace{-4pt}
\end{equation}
}
We can obtain the marginals of both features as:
\begin{equation}
\begin{aligned}
p\left(\mathbf{x}_{1}\right) &=\mathcal{N}\left(\mathbf{x}_{1} \mid \boldsymbol{\mu}_{1}, \boldsymbol{\Sigma}_{11}\right) \\
p\left(\mathbf{x}_{2}\right) &=\mathcal{N}\left(\mathbf{x}_{2} \mid \boldsymbol{\mu}_{2}, \boldsymbol{\Sigma}_{22}\right),
\end{aligned}
\label{eq:missing_marginals}
\end{equation}
which finally allows us to compute the posterior of the missing values conditioning on the observed ones:
\begin{equation}
\begin{aligned}
p\left(\mathbf{x}_{1} | \mathbf{x}_{2}\right) &=\mathcal{N}\left(\mathbf{x}_{1} | \boldsymbol{\mu}_{1 | 2}, \mathbf{\Sigma}_{1 | 2}\right),
\end{aligned}
\label{eq:missing_posterior}
\end{equation}
where
\begin{equation}
\begin{aligned}
\boldsymbol{\mu}_{1 | 2} &=\boldsymbol{\mu}_{1}+\boldsymbol{\Sigma}_{12} \boldsymbol{\Sigma}_{22}^{-1}\left(\mathbf{x}_{2}-\boldsymbol{\mu}_{2}\right) \\
&=\boldsymbol{\mu}_{1}-\boldsymbol{\Lambda}_{11}^{-1} \boldsymbol{\Lambda}_{12}\left(\mathbf{x}_{2}-\boldsymbol{\mu}_{2}\right) \\
&=\boldsymbol{\Sigma}_{1 | 2}\left(\boldsymbol{\Lambda}_{11} \boldsymbol{\mu}_{1}-\boldsymbol{\Lambda}_{12}\left(\mathbf{x}_{2}-\boldsymbol{\mu}_{2}\right)\right),
\end{aligned}
\label{eq:missing_posterior}
\end{equation}

and

\begin{equation}
\begin{aligned}
\boldsymbol{\Sigma}_{1 | 2} &=\boldsymbol{\Sigma}_{11}-\boldsymbol{\Sigma}_{12} \boldsymbol{\Sigma}_{22}^{-1} \boldsymbol{\Sigma}_{21}=\boldsymbol{\Lambda}_{11}^{-1}.
\end{aligned}
\label{eq:missing_posterior}
\end{equation}

\section{Experiments}
\label{sec:appendix_experiments}

\begin{figure}[]
\centering
\includegraphics[width=0.5\textwidth]{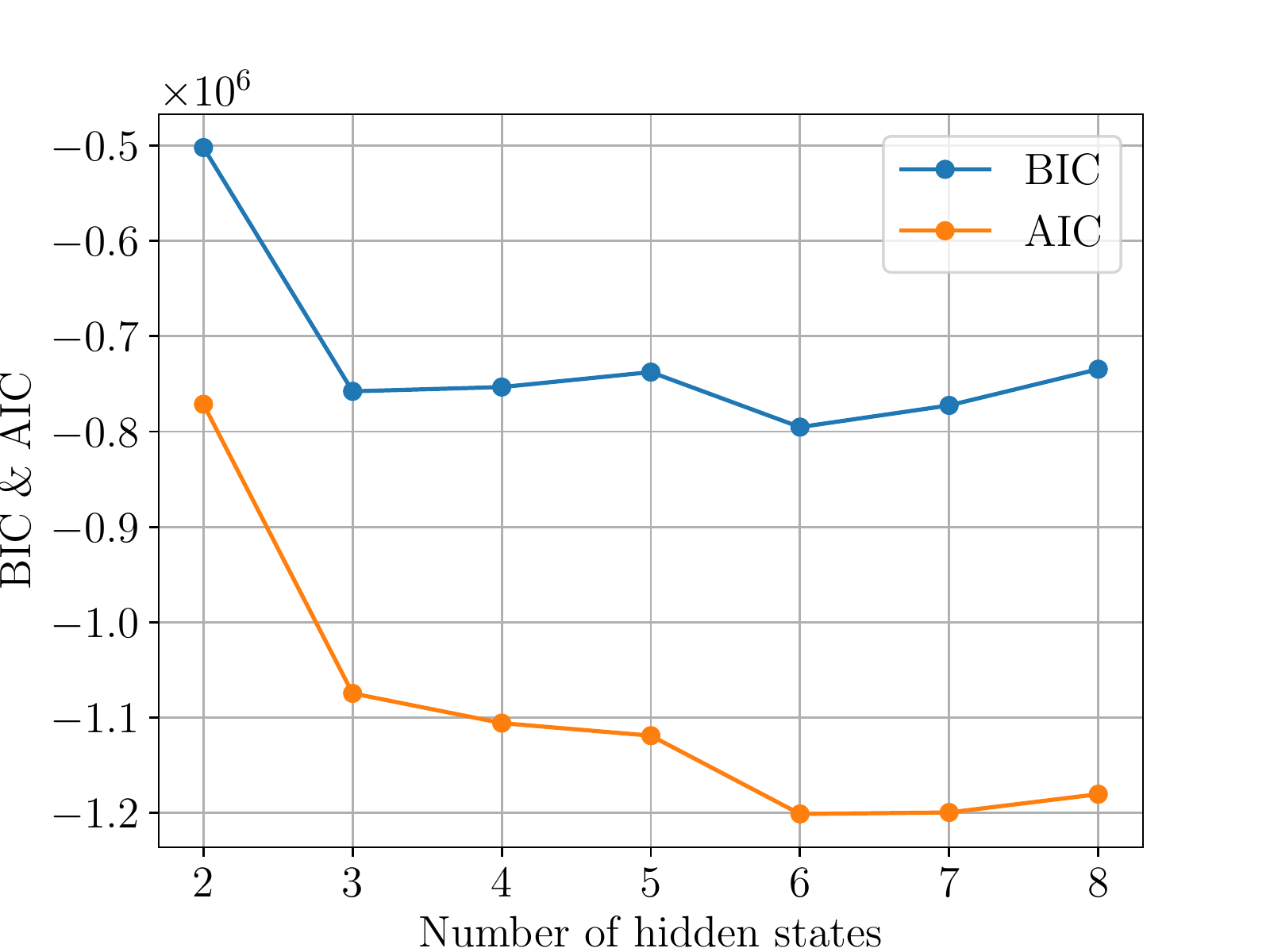}  % Reduce the figure size so that it is slightly narrower than the column.
\caption{BIC \& AIC analysis}
\label{fig:BIC}
\end{figure}

Figure \ref{fig:BIC} shows BIC and AIC evolution for different numbers of hidden states. $N=6$ was selected as the optimal value, with complexity increasing for a higher number of states. The mapping from the six hidden states to the binary classification (asleep/awake) is done by taking advantage of the capability of our proposed model to train in a semi-supervised way. Semi-supervised training within our framework allows users to manually design certain aspects of the model’s behaviour, which is especially important when training models from smartphones’ data. This allows users to fix certain discrete probability distributions. In our case, we fix the probability of sleeping to zero when observing that the patient is using the smartphone for those hidden states we aim to interpret as sleeping (while letting the model infer the observation probabilities for the rest of the features). This makes the mapping from the original six hidden states to the binary classification straightforward.

\begin{figure}[h]
\centering
    {\includegraphics[width=7.5cm, height=6.5cm]{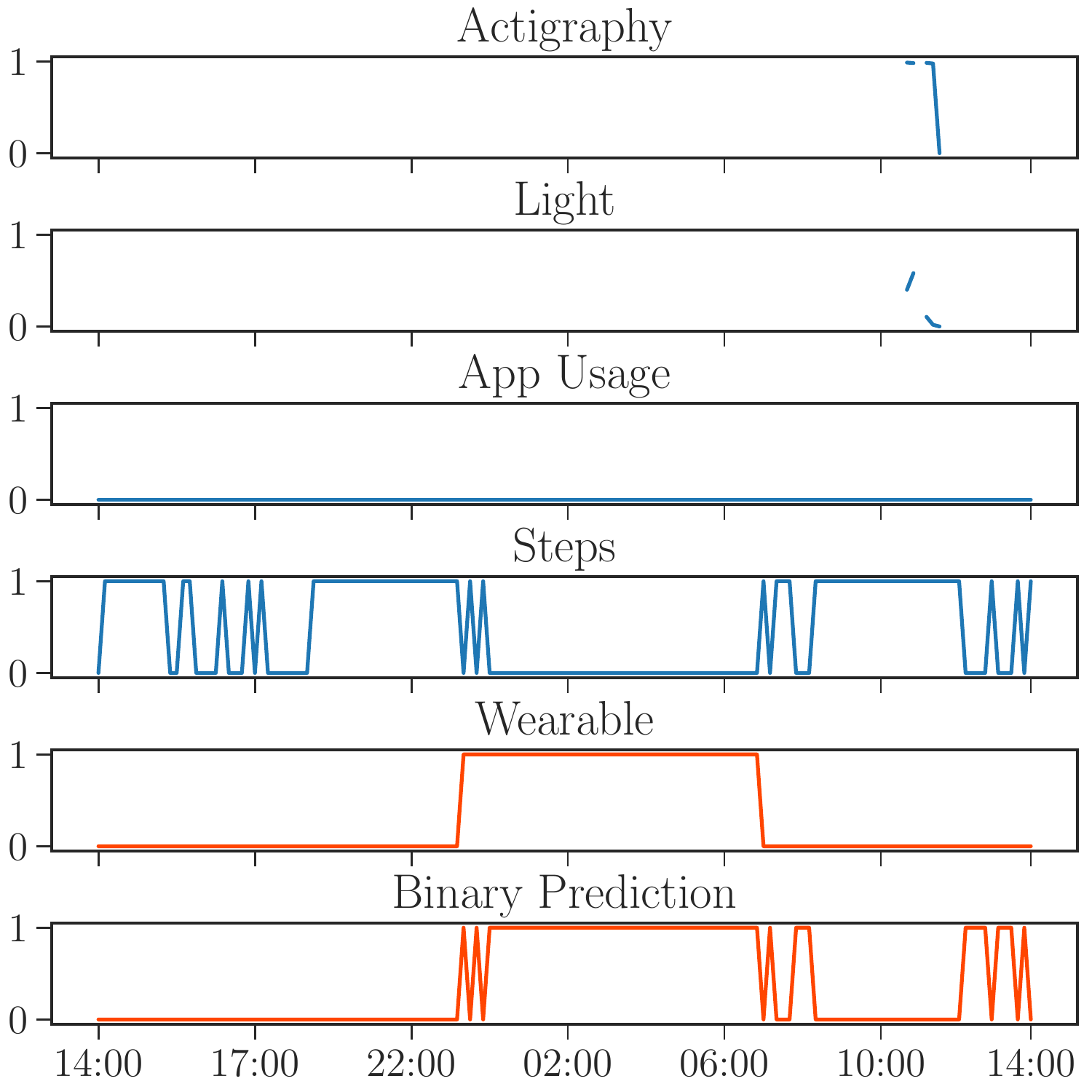}}
    %\vspace{-0.35cm}
    %\qquad
    %\subfloat[Test-set features with 10\% and 20\% abstention rate after feature selection.]{
    %{\includegraphics[width=5.0cm]{./images/plot_of_0.1sk}}
    %{\includegraphics[width=5.0cm]{./images/plot_of_0.2sk}}}
\caption{HHMM's sleep activity predictions and wearables' reported metrics comparison. Scenario with high proportion of missing data.}
\label{fig:predictions1}
%\vspace{-0.35cm}
\end{figure}

Figures \ref{fig:predictions1} and \ref{fig:predictions1}   shows two examples of the HHMM's reported predictions for different patients. The first sequence suffered from scarce missing data, gathered in the continuous features, while the second sequence exhibited significant amounts of missing values for most of the features. HHMM's prediction for the first sequence correctly integrated the available sources and the model could detect awake periods that the wearable device  missed, as the app usage feature proves. For the second sequence, the model failed predicting as sleep activity some periods distant from the primary sleep trend due to the lack of available data. This behavior could be corrected  by applying post-processing techniques to the model's predictions.

\begin{figure}[h]
\centering
    {\includegraphics[width=7.5cm, height=6.5cm]{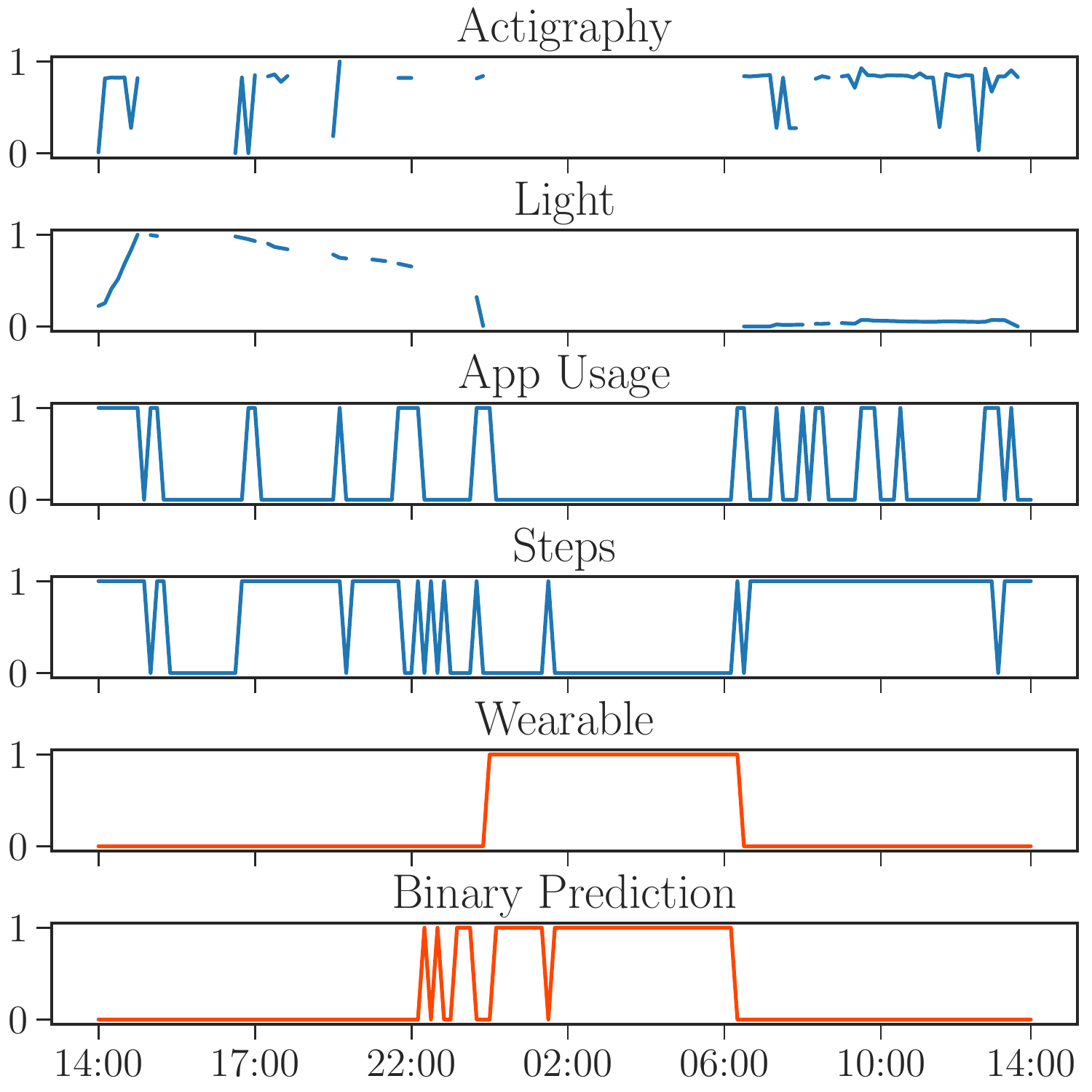}}\\
    %\vspace{-0.35cm}
    %\qquad
    %\subfloat[Test-set features with 10\% and 20\% abstention rate after feature selection.]{
    %{\includegraphics[width=5.0cm]{./images/plot_of_0.1sk}}
    %{\includegraphics[width=5.0cm]{./images/plot_of_0.2sk}}}
\caption{HHMM's sleep activity predictions and wearables' reported metrics comparison. Scenario with low proportion of missing data.}
\label{fig:predictions2}
%\vspace{-0.35cm}
\end{figure}

\end{document}